\documentclass[final,5p,twocolumn]{elsarticle}

\usepackage{amssymb}
\usepackage{amsmath}
\usepackage{xcolor}

\journal{Nuclear Physics B}

\begin{document}

\begin{frontmatter}



\title{Uncertainty-Aware Mapping from 3D Keypoints to Anatomical Landmarks for Markerless Biomechanics}


\author[label1]{Cesare Davide Pace\corref{cor1}}
\ead{cesaredavide.pace@unicas.it}
\cortext[cor1]{Corresponding author}

\author[label2,label3]{Alessandro Marco De Nunzio}

\author[label1]{Claudio De Stefano}
\author[label1]{Francesco Fontanella}
\author[label1]{Mario Molinara}

\affiliation[label1]{organization={Department of Electrical and Information Engineering, University of Cassino and Southern Lazio},
      country={Italy}}

\affiliation[label2]{organization={Department of Research and Development, LUNEX},
      country={Luxembourg}}

\affiliation[label3]{organization={Luxembourg Health \& Sport Sciences Research Institute ASBL},
      country={Luxembourg}}

\begin{abstract}
Markerless biomechanics increasingly relies on 3D skeletal keypoints extracted from video, yet downstream biomechanical mappings typically treat these estimates as deterministic, providing no principled mechanism for frame-wise quality control. In this work, we investigate predictive uncertainty as a quantitative measure of confidence for mapping 3D pose keypoints to 3D anatomical landmarks, a critical step preceding inverse kinematics and musculoskeletal analysis.

Within a temporal learning framework, we model both uncertainty arising from observation noise and uncertainty related to model limitations. Using synchronized motion capture ground truth on AMASS, we evaluate uncertainty at frame and joint level through error--uncertainty rank correlation, risk--coverage analysis, and catastrophic outlier detection.

Across experiments, uncertainty estimates—particularly those associated with model uncertainty—exhibit a strong monotonic association with landmark error (Spearman $\rho \approx 0.63$), enabling selective retention of reliable frames (error reduced to $\approx 16.8$\,mm at 10\% coverage) and accurate detection of severe failures (ROC-AUC $\approx 0.92$ for errors $>50$\,mm). Reliability ranking remains stable under controlled input degradation, including Gaussian noise and simulated missing joints. In contrast, uncertainty attributable to observation noise provides limited additional benefit in this setting, suggesting that dominant failures in keypoint-to-landmark mapping are driven primarily by model uncertainty.

Our results establish predictive uncertainty as a practical, frame-wise tool for automatic quality control in markerless biomechanical pipelines.
\end{abstract}


\end{frontmatter}

\section{Introduction}
\label{sec:introduction}

The quantitative analysis of human movement increasingly relies on intermediate representations derived from pattern recognition and machine learning techniques, such as skeletal keypoints and latent motion descriptors. These representations underpin modern markerless pipelines, enabling scalable analysis across domains including biomechanics, rehabilitation, and gesture analysis. Recent advances in computer vision have made it possible to extract 2D and 3D human pose keypoints directly from monocular or multi-view video, substantially lowering the barrier to motion analysis outside controlled laboratory environments \cite{lam2023systematic, scataglini2024accuracy}.

Marker-based motion capture remains the gold standard due to its anatomical grounding and spatial precision, yet its cost and infrastructure requirements limit scalability. \cite{d2024validation}. Markerless alternatives based on pose estimation and kinematic reconstruction offer a compelling opportunity to democratize movement analysis and enable large-scale applications \cite{uhlrich2023opencap, wade2022applications}.

Despite this progress, current markerless pipelines suffer from a fundamental limitation: intermediate pose representations are treated as deterministic quantities \cite{bramlage2023plausible}. Neural networks typically output a single point estimate for each joint or keypoint, which is then propagated through downstream modules such as anatomical landmark regression, inverse kinematics, or musculoskeletal simulation. Performance is commonly summarized using average accuracy metrics aggregated over entire sequences or datasets \cite{kappan2026survey}. While informative at a global level, these metrics fail to capture frame-wise reliability and often mask localized failures caused by occlusions, atypical postures, or rapid movements.

This limitation is particularly critical in biomechanics, where clinically meaningful interpretations may depend on precise estimates at specific movement phases. A model can exhibit low average error while producing severe outliers during critical gait events, such as toe-off or peak knee flexion. In contrast to marker-based systems, which provide implicit quality indicators through physical markers and signal integrity, markerless pipelines lack a principled mechanism to distinguish reliable anatomical estimates from hallucinated predictions driven by model uncertainty \cite{wade2022applications}. As a result, quality control is often performed manually or via ad hoc filtering, undermining scalability.

A key challenge, therefore, is not improving accuracy per se, but determining when a prediction should be trusted. This calls for reliability-aware representations that explicitly quantify predictive confidence \cite{hauenstein2024reliability}. In the broader machine learning literature, predictive uncertainty has emerged as a principled framework to address this need, commonly decomposed into aleatoric uncertainty, reflecting irreducible observation noise, and epistemic uncertainty, capturing uncertainty in the model parameters due to limited or biased training data \cite{kendall2017uncertaintiesneedbayesiandeep, zou2023review}.

However, uncertainty modeling remains largely underexplored in human movement analysis, particularly in the mapping from pose keypoints to anatomical landmarks required for biomechanical interpretation. Indeed, as mentioned above, existing approaches are predominantly deterministic and provide no mechanism to flag unreliable frames or joints, nor to support automatic quality control in downstream analyses \cite{uhlrich2023opencap, ruescas2024deep}. Reliability is therefore typically assessed retrospectively using static agreement metrics, which are ill-suited to capture the temporal and joint-specific nature of markerless failure modes.

In this work, we investigate whether predictive uncertainty can serve as a reliable, frame-wise measure of confidence in the mapping from 3D pose keypoints to 3D anatomical landmarks. Rather than focusing on architectural novelty or marginal improvements in accuracy, we analyze the relationship between uncertainty and true biomechanical error and evaluate whether uncertainty enables risk-controlled inference in markerless pipelines. We model both aleatoric and epistemic uncertainty within a temporal learning framework and assess their behavior using synchronized motion capture ground truth.

Our experiments show that epistemic uncertainty exhibits a strong, monotonic relationship with landmark estimation error across joints and over time, while aleatoric uncertainty provides limited additional benefit in this setting. These findings highlight the importance of uncertainty-aware modeling for reliable human movement analysis and motivate its use as a principled tool for automatic quality control.

The main contributions of this work can be summarized as follows:
\begin{itemize}
 \item An uncertainty-aware framework for 3D keypoint-to-landmark mapping.
 \item Frame-wise analysis of uncertainty–error correlation.
 \item Risk–coverage and outlier detection evaluation.
 \item Empirical evidence that epistemic uncertainty dominates.
\end{itemize}

This work focuses on reliability rather than accuracy and aims to provide methodological foundations for trustworthy, large-scale markerless biomechanical analysis.

\section{Related Work}
\label{related_work}

\subsection{Markerless motion capture and biomechanical mapping}
Markerless motion capture systems estimate skeletal keypoints from video and reconstruct 3D motion for biomechanical analysis \cite{lam2023systematic, kappan2026survey}. Recent frameworks, including OpenCap, have demonstrated competitive agreement with marker-based systems under controlled conditions \cite{uhlrich2023opencap, d2024validation}. 

However, evaluation is typically based on aggregate accuracy metrics computed over complete trials, which obscure transient, frame-specific failures arising from occlusions, rapid movements, or atypical postures \cite{wade2022applications}. As a result, frame-wise reliability remains largely uncharacterized in markerless pipelines.

A critical stage in these systems is the mapping from generic 3D pose keypoints to anatomical landmarks for inverse kinematics and musculoskeletal analysis. Existing landmark estimation methods rely on deterministic regression or physics-based optimization and are typically evaluated using aggregate positional or angular errors, without providing frame-wise confidence estimates \cite{ruescas2024deep, uhlrich2023opencap, horsak2023concurrent}.

\subsection{Uncertainty modeling and reliability assessment}
Predictive uncertainty refers to the model’s quantified degree of confidence in its own predictions. In machine learning, uncertainty is commonly decomposed into aleatoric uncertainty, reflecting irreducible observation noise, and epistemic uncertainty, capturing uncertainty in model parameters due to limited or biased training data \cite{kendall2017uncertaintiesneedbayesiandeep, zou2023review}. Established techniques include heteroscedastic regression, Monte Carlo dropout, deep ensembles, and variational inference \cite{gal2016dropout, maddox2019simple}.

In computer vision, uncertainty has been investigated in human pose estimation to improve robustness, calibration, and selective prediction under distribution shift \cite{bramlage2023plausible}. More recently, probabilistic formulations have been explored in gait analysis to quantify confidence in joint kinematics and identify unreliable trials or steps. For example, Donahue et al.~\cite{donahue2026embc} proposed calibrated uncertainty estimation for clinical gait analysis using probabilistic multi-view markerless pipelines, demonstrating improved trustworthiness at the step level.

However, existing approaches predominantly operate at the level of pose estimation or downstream kinematic summaries. The use of predictive uncertainty as a frame-wise reliability signal for the regression from 3D pose keypoints to biomechanically-defined anatomical landmarks remains largely unexplored. In particular, the application of uncertainty-based risk–coverage analysis and selective inference to anatomical landmark mapping within OpenSim-style pipelines has received little attention.

This work addresses this gap by explicitly modeling and evaluating predictive uncertainty in the keypoint-to-landmark mapping stage. Rather than applying uncertainty post hoc to kinematic outputs, we investigate its role as a reliability signal at the anatomical reconstruction level, enabling frame-wise quality control in markerless biomechanics.

\subsection{Problem Formulation and Temporal Mapping}
\label{sec:problem_formulation}

Let $\mathbf{x}_{1:T} = \{\mathbf{x}_t\}_{t=1}^{T}$ denote a temporal sequence of 3D pose keypoints extracted from video or motion capture data, where each frame
\begin{equation}
\mathbf{x}_t \in \mathbb{R}^{K \times 3}, \quad K = 20,
\end{equation}
represents the 3D coordinates of $K$ skeletal keypoints. Here, $K = 20$ corresponds to the number of 3D pose keypoints retained from the AMASS skeletal representation after alignment with the target biomechanical marker set.

For convenience, each frame $\mathbf{x}_t$ is flattened before being processed by the temporal model, such that $\mathbf{x}_t \in \mathbb{R}^{3K}$.

The target output is a sequence of 3D anatomical landmarks
\begin{equation}
\mathbf{y}_t \in \mathbb{R}^{L \times 3}, \quad L = 43,
\end{equation}
corresponding to anatomical marker locations compatible with biomechanical models. Here, $L = 43$ denotes the anatomical landmark configuration adopted for compatibility with downstream inverse kinematics models. The landmark set includes body markers derived from the AMASS dataset \cite{AMASS:ICCV:2019}, with additional padded markers to ensure consistent dimensionality across subjects and configurations.

Given a fixed-length input sequence with $T = 60$, the objective is to learn a temporal mapping
\begin{equation}
f_\theta : \mathbb{R}^{T \times 3K} \rightarrow \mathbb{R}^{T \times L \times 3},
\end{equation}
that predicts anatomical landmarks for every frame in a many-to-many fashion. The choice of $T = 60$ corresponds to a temporal window of one second at the dataset sampling rate (60 Hz), providing sufficient motion context for stable landmark reconstruction while maintaining computational efficiency.

We use an LSTM-based temporal model as a representative baseline for sequence-to-sequence keypoint-to-landmark mapping \cite{hochreiter1997long, uhlrich2023opencap, ruescas2024deep}. 

Given the flattened input sequence $\mathbf{x}_{1:T}$, the network produces frame-wise predictions \( \boldsymbol{\mu}_t = f_\theta(\mathbf{x}_{1:T})_t \), where $\boldsymbol{\mu}_t \in \mathbb{R}^{L \times 3}$ denotes the predicted landmark positions at frame $t$.

The model outputs predictions for all frames, enabling frame-wise reliability analysis. Beyond point predictions $\boldsymbol{\mu}_t$, the framework is extended to quantify predictive uncertainty, as described next.

\subsection{Uncertainty Modeling}

\paragraph{Aleatoric uncertainty.}
To model observation-dependent uncertainty, we adopt a heteroscedastic regression formulation \cite{kendall2017uncertaintiesneedbayesiandeep, bishop2006pattern}, allowing the predictive variance to depend on the input. In addition to the predictive mean $\boldsymbol{\mu}_t$, the network outputs an input-dependent log-variance term $\log \boldsymbol{\sigma}^2_t \in \mathbb{R}^{L \times 3}$ under a diagonal covariance assumption.

Under a Gaussian likelihood, the negative log-likelihood loss for a single frame is defined as
\begin{equation}
\mathcal{L}_{\text{ale}}(t) =
\frac{1}{2}
\sum_{l=1}^{L}
\sum_{d=1}^{3}
\left(
\frac{(y_{t,l,d} - \mu_{t,l,d})^2}{\sigma_{t,l,d}^2}
+ \log \sigma_{t,l,d}^2
\right),
\end{equation}
where $(l,d)$ index landmarks and spatial coordinates. The log-variance is clamped during training for numerical stability and the loss is averaged over valid landmarks and frames. Aleatoric uncertainty captures irreducible observation noise conditioned on the input.

\paragraph{Epistemic uncertainty.}
Epistemic uncertainty is approximated using Monte Carlo (MC) dropout \cite{kendall2017uncertaintiesneedbayesiandeep, gal2016dropout}. At inference time, dropout layers are activated and $M$ stochastic forward passes are performed, yielding predictions $\{\boldsymbol{\mu}^{(m)}_t\}_{m=1}^{M}$. The epistemic variance is estimated as the empirical variance of these samples:
\[
\boldsymbol{\sigma}^2_{\text{epi},t}
=
\frac{1}{M}
\sum_{m=1}^{M}
\left(\boldsymbol{\mu}^{(m)}_t - \bar{\boldsymbol{\mu}}_t \right)^2.
\]
In all experiments, $M=50$ samples are used.

\paragraph{Total predictive uncertainty.}
Assuming independence between aleatoric and epistemic components, total predictive variance is computed as 
\[
\boldsymbol{\sigma}^2_{\text{tot},t}
=
\boldsymbol{\sigma}^2_{\text{epi},t}
+
\boldsymbol{\sigma}^2_{\text{ale},t}.
\]
For frame-level analysis, variance is aggregated across landmarks and spatial dimensions using the mean operator to obtain a scalar uncertainty score.

\subsection{Reliability Evaluation}

To assess whether predictive uncertainty provides a meaningful reliability signal, we employ complementary evaluation metrics.

\paragraph{Error--uncertainty correlation.}
Frame-wise prediction error is defined as
\begin{equation}
e_t = \frac{1}{L}\sum_{l=1}^{L} \|\mathbf{y}_{t,l} - \boldsymbol{\mu}_{t,l}\|_2,
\end{equation}
and Spearman rank correlation is computed between $e_t$ and the corresponding uncertainty score $u_t$ to evaluate monotonic association.

\paragraph{Risk--coverage analysis.}
Frames are ranked by increasing uncertainty. For coverage level $c \in (0,1]$, risk is defined as $\text{Risk}(c) = \mathbb{E}[e_t \mid u_t \leq u_c]$, where $u_c$ retains the lowest $c$ fraction of frames. This analysis evaluates selective retention of reliable predictions.

\paragraph{Outlier detection.}
Frames with error exceeding a predefined threshold are treated as outliers. Uncertainty is used as a decision variable, and performance is quantified using ROC-AUC and PR-AUC metrics.

\paragraph{Input degradation.}
To evaluate robustness, zero-mean Gaussian noise with configurable standard deviation is injected into valid input keypoints at inference time. The resulting uncertainty behavior is analyzed to assess whether reliability ranking remains informative under degraded conditions.

\section{Experimental Setup}
\label{sec:experimental_setup}

This section describes the datasets, preprocessing steps, training configuration, and evaluation protocols used to assess the proposed uncertainty-aware framework. All experiments are conducted using a consistent setup to ensure reproducibility and controlled analysis of predictive reliability.

\subsection{Dataset and Preprocessing}
\label{sec:dataset_preprocessing}

Experiments were conducted on the AMASS dataset \cite{AMASS:ICCV:2019}, which aggregates motion capture recordings from 16 independent studies collected across different laboratories, subjects, acquisition protocols, and motion types. In total, AMASS comprises 1,176 subjects and approximately 221 hours of motion data, covering a wide spectrum of activities including walking, running, sports, and other dynamic movements. This large-scale aggregation provides substantial variability in body morphology and motion dynamics, supporting a realistic evaluation of generalization and reliability.

All available AMASS subsets were used without filtering by motion category or sequence duration. Motion sequences were segmented into fixed-length clips of $T=60$ frames (one second at 60 Hz). Data were split at the subject level to prevent identity leakage, using an 80\%/10\%/10\% split for training, validation, and testing. Importantly, the split was performed independently within each of the 16 constituent studies, ensuring proportional representation of all motion sources across training, validation, and test sets.

All coordinates were represented in metric space. Distances were internally computed in meters and converted to millimeters for reporting.

\subsection{Training Configuration}
\label{sec:training_details}

Models are trained using the AdamW optimizer with a learning rate of $1\times10^{-4}$ and a weight decay of $0.05$. Training runs for up to 200 epochs with batch size 64, and early stopping is applied based on validation loss (patience = 10). Hyperparameters were selected based on validation performance and preliminary stability analysis, and were kept fixed across deterministic and uncertainty-aware configurations for fair comparison.

When aleatoric uncertainty modeling is enabled, the training objective is

\begin{equation}
\mathcal{L} =
\mathcal{L}_{\text{NLL}}
+ 1.0\,\mathcal{L}_{\text{vel}}
+ 0.5\,\mathcal{L}_{\text{acc}}
+ 1.0\,\mathcal{L}_{\text{angle}}
+ 1.0\,\mathcal{L}_{\text{pos}},
\end{equation}

where $\mathcal{L}_{\text{NLL}}$ denotes the Gaussian negative log-likelihood, while the additional terms enforce temporal smoothness, angular consistency, and joint-wise positional accuracy. Loss weights were set to balance the relative magnitude of each component and were kept fixed across all experiments.

\subsection{Monte Carlo Dropout Inference}
\label{sec:mc_dropout_setup}

Epistemic uncertainty is estimated at inference time using MC dropout. During evaluation, dropout layers within the LSTM are re-enabled while keeping batch normalization layers in evaluation mode. A dropout rate of 0.1 is used.

For each input sequence, $M=50$ stochastic forward passes are performed. Predictive statistics are computed from the resulting sample distribution. Dropout is applied only during evaluation and does not affect training.

\subsection{Input Degradation Experiments}
\label{sec:noise_setup}

To assess robustness and uncertainty behavior under degraded inputs, controlled perturbations are applied to the input keypoints at inference time.

Gaussian noise with zero mean and standard deviations $\sigma \in \{0, 2.5, 5, 7.5, 10, 15, 20, 30\}$ mm is independently added to valid keypoints.

\subsection{Evaluation Protocol}
\label{sec:evaluation_protocol}

Prediction accuracy is measured using the Euclidean distance between predicted and ground-truth landmarks. Errors are computed per joint and averaged across landmarks to obtain a frame-wise error, which is then aggregated across frames and sequences.

Reliability is evaluated using (i) Spearman rank correlation between predicted uncertainty and true error, (ii) risk--coverage curves measuring expected error as a function of retained coverage, and (iii) outlier detection performance quantified via ROC-AUC and PR-AUC.

Note that the absolute magnitude of predicted uncertainty is not interpreted directly; instead, we focus on relative variations and ranking consistency with respect to true error.



\section{Results}
\label{sec:results}

\subsection{Baseline Landmark Estimation Accuracy}
\label{sec:baseline_accuracy}

We first reported the performance of the deterministic temporal model on the AMASS test set. Accuracy was evaluated using the mean per-joint position error (MPJPE).

The baseline model achieved a test MPJPE of 21.14~mm, confirming that the proposed temporal mapping yields accurate landmark estimates under nominal conditions. For reference, the average spatial range of anatomical landmark trajectories in the test set was approximately 279 mm (computed over 20,329 sequences), indicating that the absolute error corresponds to roughly 7.6\% of the typical movement amplitude.

We then evaluated a heteroscedastic variant of the same model, trained to predict both landmark means and input-dependent variance. To ensure stable optimization, the heteroscedastic model was initialized from the pretrained deterministic baseline and fine-tuned using a likelihood-based objective, as heteroscedastic regression is known to be sensitive to initialization.

Using this warm-start strategy, the heteroscedastic model achieved a test MPJPE of 25.51~mm ($\approx 9.1\%$ of the average movement range). Although point-wise accuracy degraded compared to the deterministic baseline, performance remained within a reasonable range, enabling meaningful analysis of predictive uncertainty. Training the heteroscedastic model from scratch consistently yielded substantially lower accuracy. 

\subsection{Quantitative Validation of Uncertainty as a Reliability Signal}
\label{sec:uncertainty_error}
We analyze the relationship between predictive uncertainty and true landmark estimation error at the frame level. Uncertainty estimates are obtained using the heteroscedastic model augmented with MC dropout at inference time. Spearman rank correlation is used to quantify the monotonic relationship between frame-wise uncertainty and error.

Epistemic uncertainty exhibited a strong correlation with true landmark error ($\rho = 0.63$), indicating that higher uncertainty values are consistently associated with larger biomechanical errors. In contrast, aleatoric uncertainty (not shown) yielded substantially weaker correlations, suggesting that reliability in this task is primarily driven by model uncertainty.

\begin{figure}[t]
 \centering
 \includegraphics[width=0.80\linewidth]{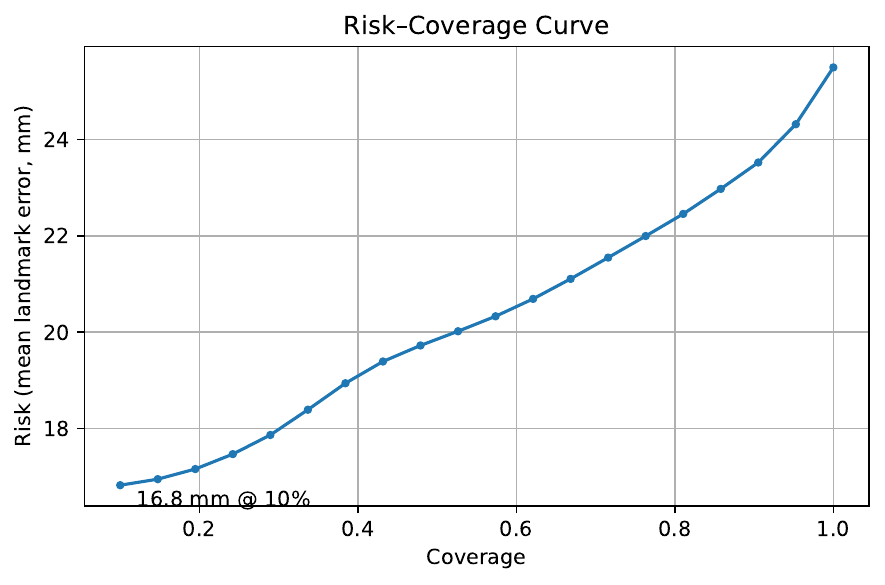}
 \caption{Risk--coverage curve obtained by ranking frames according to epistemic uncertainty. The smooth increase in risk indicates a well-ordered ranking of frame-wise reliability, enabling effective uncertainty-guided filtering.}
 \label{fig:risk_coverage}
\end{figure}

To evaluate practical utility, we performed a risk--coverage analysis by ranking frames according to epistemic uncertainty. Figure~\ref{fig:risk_coverage} showed a smooth and approximately linear increase in risk as coverage increased, indicating a well-ordered ranking of frame-wise reliability. Retaining only the 10\% most confident frames reduced the expected landmark error to 16.8 mm ($\approx 6.0\%$ of the typical movement range), corresponding to a 34\% relative reduction, demonstrating the practical utility of uncertainty-guided selection.

We further evaluate uncertainty for detecting catastrophic failures, defining outliers as frames with mean landmark error exceeding 50~mm. Epistemic uncertainty achieved strong outlier detection performance (ROC-AUC = 0.92, PR-AUC = 0.39), demonstrating its effectiveness at identifying rare but severe failure cases.

We additionally investigated whether combining aleatoric and epistemic uncertainty provides complementary reliability information. We found that heteroscedastic regression combined with MC dropout does not improve reliability metrics compared to epistemic uncertainty alone: Spearman correlation, risk--coverage behavior, and outlier detection performance remain virtually unchanged. Total uncertainty closely overlapped with epistemic uncertainty across all metrics.

These results indicated that failure cases in keypoint-to-landmark mapping are primarily driven by model uncertainty rather than input-dependent observation noise.

\subsection{Robustness of the Reliability Ranking under Degradation}
\label{sec:robustness}

We evaluated the stability of the uncertainty-based reliability ranking under controlled input degradation by injecting zero-mean Gaussian noise of increasing magnitude into the 3D input keypoints at inference time. This stress test, conducted on the validation split, assessed whether uncertainty preserved its ordering capability under progressively corrupted observations.

\begin{table}[t]
\centering
\caption{Robustness under Gaussian noise injection on the validation split.}
\label{tab:noise_robustness}
\begin{tabular}{lccc}
\hline
$\sigma$ (mm) & Mean error (mm) & Spearman $\rho$ & ROC-AUC \\
\hline
0.0 & 25.70 & 0.592 & 0.899 \\
5.0 & 25.81 & 0.592 & 0.899 \\
10.0 & 26.13 & 0.594 & 0.899 \\
20.0 & 27.39 & 0.597 & 0.898 \\
30.0 & 29.45 & 0.593 & 0.896 \\
\hline
\end{tabular}
\end{table}

As noise increases from $\sigma=0$ to $30$~mm, landmark position and joint angle errors increased monotonically (we report representative noise levels in Table~\ref{tab:noise_robustness}), with mean error rising from 25.70~mm to 29.45~mm. Despite this substantial degradation in absolute accuracy, the uncertainty--error relationship remains remarkably stable: Spearman correlation consistently stayed within $\rho=0.592$--$0.597$, indicating that the relative ranking of reliable versus unreliable frames is largely unaffected by noise. Risk--coverage curves shifted upward due to increased error magnitude, yet preserve their shape, confirming that uncertainty retains its discriminative ordering.

Outlier detection performance remained high across all noise levels (ROC-AUC $=0.899$ at $\sigma=0$ and $0.896$ at $\sigma=30$~mm), showing only marginal variation. This minimal decrease further demonstrates that epistemic uncertainty maintains strong separation between nominal and catastrophic frames even under severe perturbations.

These results provided quantitative evidence that epistemic uncertainty acts as a robust reliability signal, preserving frame-wise ranking consistency even when input keypoints are substantially corrupted.

\section{Discussion}
\label{sec:discussion}

This study examined predictive uncertainty as a reliability signal for mapping 3D pose keypoints to anatomical landmarks in markerless biomechanics. While our experiments rely on an LSTM-based temporal mapping model, chosen as a representative and well-established solution in this domain, the focus is not on architectural novelty but on whether uncertainty estimates meaningfully reflect true biomechanical error and can support frame-wise quality control in practical analysis pipelines.

Across all experiments, epistemic uncertainty consistently emerged as the most informative component. It showed a strong monotonic relationship with landmark estimation error, enabled effective risk-coverage trade-offs, and reliably detected severe failure cases. Importantly, these properties remained stable under controlled input degradation, including Gaussian noise injection. This demonstrates that epistemic uncertainty provides a robust ranking of frame-wise reliability, rather than a brittle or overconfident confidence score.

From a biomechanical perspective, these results are consistent with common failure modes observed in markerless motion analysis. Large landmark errors often arise during underrepresented or ambiguous motion configurations, such as extreme joint flexion, self-occlusions, or rapid transitions between movement phases. In such cases, downstream biomechanical computations are particularly sensitive to small geometric errors. The strong association between epistemic uncertainty and landmark error indicates that uncertainty can serve as an early indicator of these biomechanically critical failure modes, even when average accuracy remains acceptable.

In contrast, aleatoric uncertainty provided limited additional benefit beyond epistemic uncertainty. Combining heteroscedastic regression with MC Dropout did not improve correlation, selective inference, or outlier detection performance. This finding should be interpreted in light of the experimental setting: inputs consist of reconstructed 3D keypoints with relatively limited observation noise, and dominant errors are driven by model generalization rather than sensor-level variability. In this context, uncertainty primarily reflects epistemic factors such as insufficient training coverage of specific postures or motion dynamics. Notably, the reliability metrics and uncertainty–error analyses considered in this work are model-agnostic and can, in principle, be applied to alternative mappings such as feed-forward or Transformer-based temporal architectures.

A limitation of this study is that uncertainty behavior was evaluated using a single temporal architecture (LSTM). This choice was intentional, as our objective is not architectural comparison but to assess whether predictive uncertainty provides a meaningful reliability signal in keypoint-to-landmark mapping. The uncertainty mechanisms considered here are architecture-agnostic and can be integrated into alternative sequence models.

These findings have direct implications for decision-making in markerless biomechanics. Rather than treating all frames equally, uncertainty estimates can be used to automatically exclude unreliable segments and weight observations in downstream analyses. In clinical or large-scale research settings, this enables a shift from manual quality control to principled, uncertainty-guided selection of reliable data, improving both scalability and reproducibility.

\section{Conclusion}
\label{sec:conclusion}

We presented an uncertainty-aware framework for mapping 3D pose keypoints to 3D anatomical landmarks and systematically evaluated predictive uncertainty as a measure of reliability in markerless biomechanics. Our results show that epistemic uncertainty strongly correlates with true biomechanical error, supports effective risk-controlled inference, and reliably identifies catastrophic failures.

Modeling aleatoric uncertainty through heteroscedastic regression provided limited additional benefit beyond epistemic uncertainty, indicating that failure modes in this task are predominantly driven by model uncertainty rather than observation noise. These findings suggest that epistemic uncertainty alone may be sufficient for practical quality control in keypoint-to-landmark mapping.

While evaluated using a single temporal architecture, the proposed framework is not model-specific. Future work will investigate uncertainty behavior across alternative architectures, such as Transformer-based models, and extend the analysis to end-to-end markerless pipelines to study uncertainty propagation across stages.

This work establishes predictive uncertainty as a principled and practical reliability signal for markerless motion analysis and provides a foundation for uncertainty-aware quality control in large-scale biomechanical pipelines.

\section*{Declaration of generative AI and AI-assisted technologies in the manuscript preparation process}

During the preparation of this manuscript, the authors used ChatGPT in order to improve language and readability. The authors reviewed and take full responsibility for the final content.

\bibliographystyle{elsarticle-num-names} 
\bibliography{bib.bib}

\end{document}